\documentclass[11pt,reqno]{article}

\usepackage{amsmath}
\usepackage{amsfonts}
\usepackage{amssymb}
\usepackage{amsxtra}
\usepackage{graphicx}
\usepackage{caption}
\usepackage{ushort}
\usepackage{color}
\usepackage{hyperref}
\usepackage[parsep]{collref}
\hypersetup{linktocpage=true}

 \def\bd{\begin{document}} \def\ed{\end{document}}
\def\ds{\documentstyle} \let\fr=\frac \let\bl=\bigl \let\br=\bigr
\let\Br=\Bigr \let\Bl=\Bigl
\let\bm=\bibitem
\let\na=\nabla
\let\pa=\partial \let\ov=\overline

\newcommand{\be}{\begin{equation}}
\newcommand{\ee}{\end{equation}}
\newcommand{\bea}{\begin{eqnarray}}
\newcommand{\eea}{\end{eqnarray}}
\newcommand{\ba}{\begin{array}}
\newcommand{\ea}{\end{array}}

\def\ft#1#2{{\textstyle{{\scriptstyle #1}\over {\scriptstyle #2}}}}
\def\fft#1#2{{#1 \over #2}}\def\del{\partial}
\def\vp{\varphi}
\def\sst#1{{\scriptscriptstyle #1}}
\def\st#1{{\scriptstyle #1}}
\def\oneone{\rlap 1\mkern4mu{\rm l}}
\def\td{\tilde}
\def\wtd{\widetilde}
\def\ie{{\it i.e.\ }}
\def\iec{{\it i.e.,\ }}
\def\eg{{\it e.g.\ }}
\def\egc{{\it e.g.,\ }}
\def\dalemb#1#2{{\vbox{\hrule height .#2pt
        \hbox{\vrule width.#2pt height#1pt \kern#1pt
                \vrule width.#2pt}
        \hrule height.#2pt}}}
\def\smsquare{\mathord{\dalemb{6.8}{7}\hbox{\hskip1pt}}}
\newcommand{\ho}[1]{$\, ^{#1}$}
\newcommand{\hoch}[1]{$\, ^{#1}$}
\newcommand{\ra}{\rightarrow}
\newcommand{\lra}{\longrightarrow}
\newcommand{\Lra}{\Leftrightarrow}
\newcommand{\ap}{\alpha^\prime}
\newcommand{\bp}{\tilde \beta^\prime}
\newcommand{\tr}{{\rm tr} }
\newcommand{\Tr}{{\rm Tr} }
\def\0{{\sst{(0)}}}
\def\1{{\sst{(1)}}}
\def\2{{\sst{(2)}}}
\def\3{{\sst{(3)}}}
\def\4{{\sst{(4)}}}
\def\5{{\sst{(5)}}}
\def\6{{\sst{(6)}}}
\def\7{{\sst{(7)}}}
\def\8{{\sst{(8)}}}
\def\9{{\sst{(9)}}}
\def\ten{{\sst{(10)}}}
\def\n{{\sst{(n)}}}
\def\cA{{{\cal A}}}
\def\cF{{{\cal F}}}
\def\tV{\widetilde V}
\def\tW{\widetilde W}
\def\tH{\widetilde H}
\def\tE{\widetilde E}
\def\tF{\widetilde F}
\def\tA{\widetilde A}
\def\im{{{\rm i}}}
\def\tY{{{\wtd Y}}}
\def\ep{{\epsilon}}
\def\vep{{\varepsilon}}
\def\bD{{{\bar D}}}
\def\alp{{{\a'}^3}}
\def\bD{{{\bar D}}}
\def\R{{{\mathbb R}}}
\def\C{{{\mathbb C}}}
\def\E{{{\mathbb E}}}
\def\H{{{\mathbb H}}}
\def\CP{{{\mathbb C}{\mathbb P}}}
\def\RP{{{\mathbb R}{\mathbb P}}}
\def\Z{{{\mathbb Z}}}
\def\bA{{{\mathbb A}}}
\def\bB{{{\mathbb B}}}
\def\bC{{{\mathbb C}}}
\def\bR{{{\mathbb R}}}
\def\bD{{{\mathbb D}}}
\def\bE{{{\mathbb E}}}
\def\bZ{{{\mathbb Z}}}
\def\Re{{{\frak{Re}}}}
\def\Im{{{\frak{Im}}}}
\def\cosec{{\,\hbox{cosec}\,}}
\def\Gm{{\Gamma_{\!\! -}}}
\def\Gp{{\Gamma_{\!\! +}}}

\def\cosech{{\hbox{cosech}}}
\def\sech{{\hbox{sech}}}

\newcommand{\caltech}{\it Walter Burke Institute for Theoretical Physics, California Institute of Technology, Pasadena, CA 91125}

\newcommand{\brandeis}{\it Physics Department, Brandeis University, Waltham, MA 02454}

\newcommand{\imperial}{\it The Blackett Laboratory, Imperial College London\\
Prince Consort Road, London SW7 2AZ}

\newcommand{\auth}{
S. Deser\,\footnote{\,email: deser@brandeis.edu}\hoch{\star,\dagger},
and K.S. Stelle\,\footnote{\,email: k.stelle@imperial.ac.uk}\hoch{\ddagger}} 

\let\oldabstract\abstract
\let\oldendabstract\endabstract
\makeatletter
\renewenvironment{abstract}
{\renewenvironment{quotation}%
               {\list{}{\addtolength{\leftmargin}{1em} 
                        \listparindent 1.5em%
                        \itemindent    \listparindent%
                        \rightmargin   \leftmargin%
                        \parsep        \z@ \@plus\p@}%
                \item\relax}%
               {\endlist}%
\oldabstract}
{\oldendabstract}
\makeatother

\begin{document}
\setcounter{page}{0}
\thispagestyle{empty}
\begin{flushright}
\hfill{
BRX-TH 6655\\
CALT-TH  2019-027\\ 
Imperial/TP/2019/KSS/01}\\
\end{flushright} 

\begin{center}  

{\Large {\bf Field redefinition's help in constructing non-abelian gauge theories}}   

\vspace{15pt}

\auth

\vspace{7pt}
{\hoch{\star}\caltech} 

\vspace{7pt}
{\hoch{\dagger}\brandeis}

\vspace{7pt}
{\hoch{\ddagger}\imperial}

\end{center}

\vspace{10pt}

\begin{abstract}

We study, using the example of general covariance, to what extent a would-be non-abelian extension of free field abelian gauge theory can be helped by a field redefinition; answer -- not much! However, models resulting from dimensional reduction also include non-gauge fields needing to be integrated out, thereby offering a wider choice of redefinitions whose effects may indeed change the situation.
\end{abstract}
\pagenumbering{arabic}
\setcounter{page}{1}
\setcounter{footnote}{0}
 
\vspace{1cm}

A pervasive feature in attempts to construct nonabelian gauge theories that are ultimately seen to be inconsistent is that the first -- abelian invariant quadratic -- action term exists, as does the next, cubic one, taken as the product ``$J^\mu A_\mu$'' of the (quadratic) conserved abelian invariant current and the putative gauge field, hence also abelian invariant. This encouraging start masks the fact that the peril lies in the next, quartic, order. Indeed, many higher-spin interacting models have foundered here, not realizing that the cubic level is trivial. As an explicit familiar example, the $SU(2)$ Yang-Mills cubic action term is 
\be
A = - 1/4 \int[{(\partial_\mu A_\nu -\partial_\nu A_\mu) \times A^\mu} \cdot A^\nu]= \int J^\nu \cdot A_\nu\,;\label{cubicterm}
\ee
it is obviously invariant under the abelian, Maxwellian, gauge transformations ($A_\mu \to \partial_\mu S$) valid for the quadratic part, since $\partial_\mu J^\mu=0$ on linear shell. The critical term is the quartic, here  $\int(A \times A)^2$ in $g=1$ units, because it is the first that must fulfill a  non-abelian gauge invariance requirement, including the correct coefficient. In this note, we study a concrete and important case: a would-be Einstein action, say from (improper) dimensional reduction, where exactly this occurs; we will show that the action cannot generally be made consistent by the only valid procedure -- local field redefinition -- to reinstate the correct quartic term in its expansion. 

The Einstein action $A= \int d^n x \sqrt{-g}\, R$ in any $n>2$ dimension has a unique power series expansion about flat (or indeed, any consistent, Ricci-flat) background space, depending only on the chosen metric form: covariant, contravariant or some density version thereof. Once this convention is adopted, say $h_{\mu\nu}=g_{\mu\nu} - \eta_{\mu\nu}$, the form of each power in $h$ is fixed, so it must be matched by any would-be candidate, up to field redefinitions -- the only freedom and the one we study here. We begin by closing a couple of blind alleys. First, no improvement is possible simply by changing conventions, say by going from the co-- to the contra--variant metric expansion. This is obvious from the uniqueness of any expansion: if one expansion doesn't work, neither can any other sum to $\int \sqrt{-g}\,R$. The second is the uselessness of non-local field redefinitions with their associated new degree of freedom problems and in any case their ineffectiveness because the problems with new degrees of freedom proliferate to ever higher orders. Start with the linear action
\be
A(2) =1/2 \int d^n x\, h_{\mu\nu} O^{\mu\nu\rho\sigma} h_{\rho\sigma} =\int d^n x\, h_{\mu\nu} G^{\mu\nu}\hbox{(lin)}\,,\label{linact}
\ee
and shift $h_{\mu\nu}$ by $D^{\mu\nu\rho\sigma} C_{\rho\sigma}$, where $D$ is the (nonlocal) propagator inverse to the linear Einstein operator $O$ and $C$ is the cubic in $h$ coefficient of $h_{\mu\nu}$ in the ``bad'' quartic. This indeed would remove the latter but the price is a nonlocal fifth power term from the resulting shift in the cubic part of R. At every step a further unacceptable non-locality would be introduced. 

The remaining means to remove the quartic ``deficit'' is local field redefinition. In general this deficit, the difference between the existing quartic and the correct, Einstein one, suffers both in having the wrong overall scale factor and the wrong combination of the $h^4$ monomials. Reverse engineering easily tell us what modifications are permitted at order $h^4$, forgetting their higher power effects. Since all terms in the expansion are of the same, second derivative, order, useful field redefinitions must be algebraic, here $h_{\mu\nu} \to h_{\mu\nu} + (h^3)_{\mu\nu}$. [There can obviously be no $(h^2)_{\mu\nu}$  redefinitions because they would disturb the (assumed trivially correct) cubic terms.] The resulting quartic modification is 
\be
\Delta A(4)\sim \int d^n x\, h_{\mu\nu} O^{\mu\nu\rho\sigma} (h^3)_{\rho\sigma} = \int d^n x\, G^{\rho\sigma}\hbox{(lin)} (h^3)_{\rho\sigma}\,,\label{quartmod}
\ee
not a very general form, even allowing for integrations by parts in the $\partial^2$ structure $O^{\mu\nu\rho\sigma}$ operator. Thus IF and only IF the culprit part of the $h^4$ term in A can be put in this manifestly ``field-redefinable'' form is there a hope of success, though even that is rather unlikely given the quintic effects of this redefinition; at best there would be an infinite series of higher power redefinitions required. For our purposes, focusing on the first dangerous -- quartic -- deviation, the condition \eqref{quartmod} already suffices to rule out most candidates. At that, GR is the most favorable case because all terms are of the same derivative order, while models such as YM are of finite number and decreasing derivative order, so obviously even less amenable to field redefinitions, that we have seen start at second derivative order due to the quadratic kinematical term.\footnote{Gravitational field redefinitions were introduced, in a different context, by G. 'tHooft and M. Veltman \cite{tHooft:1974toh}. A recent list of some of the literature on field redefinitions in effective theories may be found in \cite{Criado:2018sdb}.}

Our main take-home point is that the key test of nonabelian structure and local symmetry occurs at the fourth order in fluctuation fields, be this in a gauge theory such as Yang-Mills or in a gravitational theory. In the full analysis of a complicated system such as dimensional reduction on a manifold without Killing symmetries, preservation of lower-dimensional local symmetry, and consistency with the anticipated realization of such symmetry is to be expected only after carefully integrating out heavy (non-zero-mode) fields. Alternatively, field redefinitions including massive non-gravitational fields could be made prior to integrating them out, but these would have to prepare the eventual massive mode integrations for a structure in which they made no changes in the pure gravitational part of the theory. In either case, the crucial task becomes how to obtain the correct anticipated pure gravitational structure at fourth order in fluctuation fields after all heavy fields are integrated out.

An example of a system which initially appears to generate just such problems is reduction of type IIA supergravity on a non-compact ${\cal H}^{(2,2)}$ space which nonetheless yields an effective lower dimensional theory as a result of a mass gap in the spectrum of the corresponding transverse wave function $\xi$ \cite{Crampton:2014hia}.  Expansion of the corresponding effective action initially reveals just such difficulties at fourth order in the lower dimensional gravitational $h_{\mu\nu}$. Moreover, similar difficulties at fourth order can be encountered in a toy model variant of ordinary dimensional reduction of $D=5$ GR where instead of an extra dimensional circle one reduces on a line interval with mixed boundary conditions for the transverse wave function: Dirichlet $\xi(0)=0$ on one side and Robin $\xi^\prime(1)-\xi(1)=0$ on the other.\footnote{The various problems involving technically inconsistent dimensional reduction are a large topic. Older literature on problems of consistent and technically inconsistent Kaluza-Klein reductions can be found in Refs \cite{Duff:1984hn, Duff:1985jd, Duff:1989cr}. Details of the effective theory resulting from the ${\cal H}^{(2,2)}$ reduction of type IIA supergravity and of the mixed Dirichlet-Robin reduction of $D=5$ GR will be given in \cite{EHLS}. }

\section*{Acknowledgments} 

The work of SD was supported by the U.S. Department of Energy, Office of Science, Office of High Energy Physics, under Award Number 
de-sc0011632. The work of KSS was supported by the STFC under Consolidated Grant ST/P000762/1. KSS would like to thank the Walter Burke Institute at Caltech and the Albert Einstein Institute in Potsdam for hospitality.

\enddocument